\documentclass[%
 reprint,
amsmath,amssymb,
aps,
floatfix
]{revtex4-2}

\usepackage[utf8]{inputenc} 
\usepackage[T1]{fontenc}    
\usepackage{graphicx}       
\usepackage{dcolumn}        
\usepackage{bm}             
\usepackage{braket}         
\usepackage{booktabs}       
\usepackage{here}           
\usepackage[colorlinks=true, allcolors=blue]{hyperref} 
\usepackage{color}          
\usepackage{textcomp}       
\usepackage{comment}        
\usepackage{tabularx}       
\usepackage{mathptmx}
\usepackage{mathtools}
\usepackage{comment}

\begin{document}


\title{Description of nucleon elastic scattering off $^6$Li with the four-body continuum-discretized coupled-channels method
}

\author{Kazuyuki Ogata}
\email[]{ogata.kazuyuki.169@m.kyushu-u.ac.jp}
\affiliation{Department of Physics, Kyushu University, Fukuoka 819-0395, Japan}

\author{Shoya Ogawa}
\affiliation{Department of Physics, Kyushu University, Fukuoka 819-0395, Japan}

\date{\today}
             

\begin{abstract}
 \noindent
 {\bf Background}:
 Neutron reactions off lithium isotopes up to 50~MeV are important for nuclear data science, around the International Fusion Material Irradiation Facility (IFMIF) facility in particular.
 \\
 {\bf Purpose}:
 We aim at constructing a semi-microscopic reaction model that describes neutron elastic scattering off $^6$Li up to 50~MeV taking the breakup channels of $^6$Li into account.
 \\
 {\bf Methods}:
 We adopt the continuum-discretized coupled-channels method (CDCC) with an $\alpha+p+n$ three-body model of $^6$Li. We employ the $g$-matrix effective interaction by Jeukenne, Lejeune, and Mahaux (JLM). The renormalization factors of the real and imaginary parts of the JLM interaction are treated as free parameters.
 \\
 {\bf Results}:
 The renormalization parameter of the real part of the JLM interaction is found to be constant ($=1.1$), whereas that for the imaginary part has a smooth energy dependence. The four-body CDCC calculation with these parameters well describes the angular distributions of both proton and neutron elastic scatterings as well as the neutron total cross section and proton total reaction cross section. The applicable energy range is found to be from 7~MeV to 50~MeV.
 \\
 {\bf Conclusions}:
We have constructed a reliable reaction model for describing nucleon-$^6$Li scattering between 7~MeV and 50~MeV. This model can directly be applied to inelastic scattering and breakup reactions for $^6$Li with the help of the complex scaling method.

\end{abstract}

\maketitle

\section{introduction}\label{sec:intro}

In the deuteron-triton fusion reactor, lithium isotopes will be used as a {\lq\lq}blanket'' to decelerate neutrons and to generate tritons to be used as fuel in the reactor. The description of the reaction between fast neutron of 14.1~MeV and $^{6,7}$Li is crucially important. As an ITER blanket project, International Fusion
Material Irradiation Facility (IFMIF)~\cite{Mas04} is designed to reproduce an intensive neutron beam by using 40~MeV deuteron with a lithium target. Although the key energy of neutron at IFMIF is 14.1~MeV, the energy of the produced neutron has a wide energy spectrum up to about 50~MeV. Thus, neutron reaction data on lithium up to 50~MeV will be important for understanding the particle production by neutrons at IFMIF. Although the latest nuclear database FENDL~\cite{Fen24} covers such data up to 150~MeV, because of the lack of the high-energy neutron data, it will be still important to make a neutron reaction model off lithium targets covering the neutron energies from a few MeV to 50~MeV.

The description of nuclear reactions from a fully microscopic view, i.e., {\it ab initio} approach, is rapidly developing in recent years~\cite{Bar13,Hup14,Nav16,Gys24}. However, the application of such method at energies above 20~MeV is still challenging because of the huge number of channels of the reaction system at higher energies. The microscopic optical model approaches based on the multiple scattering theory (MST)~\cite{Fol45,Wat53,Ker59}, it is sometimes called {\it ab initio}, are also developing and have been successful in the {\lq\lq}prediction'' of nucleon-nucleus scattering data~\cite{Amo00,Toy13,Toy15,Bur19,Bur20,Vor24}. The MST has, however, a limitation of the scattering energy; at low energies, the boundary condition assumed in the MST will become inadequate. Therefore, an alternative approach will be necessary to develop a neutron reaction database mentioned above.

In Ref.~\cite{Mat11}, Matsumoto and collaborators proposed a semi-microscopic model for the neutron-$^6$Li ($n$-$^6$Li) reaction at around 10--20~MeV; $^6$Li was described with an $\alpha+d$ two-body model and the continuum-discretized coupled-channels method (CDCC)~\cite{Kam86,Aus87,Yah12} was applied to describe neutron elastic and inelastic scatterings. The key ingredient was the nucleon-nucleon ($NN$) effective interaction and they employed the $g$-matrix interaction by Jeukenne, Lejeune, and Mahaux~\cite{Jeu77} (JLM), and the renormalization parameters in the JLM interaction were determined to reproduce the reaction observables. This approach is a variant of the MST-based microscopic reaction model, which would not be applicable to the low-energy scattering. However, allowing the use of the free parameters, at least phenomenologically, the calculation well reproduced the $n$-$^6$Li reaction observables. It was discussed, however, that the model severely undershoots the $n$-$^6$Li total cross section above 25~MeV.

In the present study, we basically follow Ref.~\cite{Mat11} but with an $\alpha+p+n$ three-body model for $^6$Li. Thus, we describe the $n$-$^6$Li scattering with four-body CDCC~\cite{Mat04}. As another advantage of the present calculation, we include the closed channels, which were disregarded in Ref.~\cite{Mat11}; the definition of the closed channels will be given in Sec.~\ref{sec:form}. Lastly, we carefully investigate the energy dependence of the renormalization parameters of the JLM interaction. We apply four-body CDCC with the JLM interaction to both $n$-$^6$Li and $p$-$^6$Li scattering up to 50~MeV and compare the results with experimental data. We restrict ourselves to the elastic scattering (and total and total-reaction cross sections) because it is the most important basis of the study of the $n$-$^6$Li reaction. 

This paper is constructed as follows. In Sec.~\ref{sec:form}, we briefly overview the (semi-)microscopic CDCC formalism with many-body wave function of $^6$Li. In Sec.~\ref{sec:num}, we explain the structural model inputs and the setting of the reaction calculation. We also give a few words on the JLM interaction. We show the $n$-$^6$Li and $p$-$^6$Li cross sections in Sec.~\ref{sec:res} and compare them with experimental data. The energy dependence of the renormalization factors of the JLM interaction is also discussed. Finally, we give a summary and future perspectives in Sec.~\ref{sec:sum}.

\section{Formalism}
\label{sec:form}
In CDCC, the wave function $\Psi_{JM_J}$ of the nucleon-$^6$Li  ($N$-$^6$Li) reaction system, where $J$ and $M_J$ are the total angular momentum and its third component, respectively, is expanded in terms of eigenstates $\{ \Phi_{nIm_{I}}  \}$ of $^6$Li as
\begin{align}
\Psi_{JM_J}\left(  \boldsymbol{R},\xi\right)   &
=
\sum_{nIL}
\left[
\chi_{nIL}^{\left(J\right)}\left({R}\right) \otimes
\Phi_{nI}\left(\xi\right)
\right]_{JM_J}.
\label{3psijm}%
\end{align}
Here, $I$ is the total spin of $^6$Li, $\xi$ represents the internal coordinates, and $n$ is the energy index. $\Phi_{nIm_{I}}$ satisfies
\begin{equation}
h\Phi_{nIm_{I}}\left(  \xi\right)  =\epsilon_{nI}\Phi_{nIm_{I}%
}\left(  \xi\right),
\label{3bep}
\end{equation}
where $h$ is the Hamiltonian of $^6$Li, $\epsilon_{nI}$ is the eigenenergy, and $m_I$ is the third component of $I$.
In Eq.~\eqref{3psijm}, $\chi_{nIL}^{\left(J\right)}$ is the expansion {\lq\lq}coefficient'' with $R$ ($L$) being the distance (orbital angular momentum) between $N$ and $^6$Li. 
We use the Gaussian expansion method (GEM)~\cite{Hiy03} to solve the eigenvalue problem of Eq.~\eqref{3bep}. We employ Gaussian basis that sufficiently covers the space needed for describing the reaction process. In other words, the eigenstates above the particle threshold can be regarded as discretized continuum states of $^6$Li.

The coupled-channel (CC) equations are given by
\begin{align}
\left[  -\frac{\hbar^{2}}{2\mu}\frac{d^{2}}{dR^{2}}+\frac{\hbar^{2}}{2\mu}
\frac{L(L+1)}{R^{2}}-E_{n I}\right] \chi_{c}^{\left(  J\right)  }\left(
K_{nI},R\right) \nonumber \\
=-\sum_{c'}U^{(J)}_{cc'}\left(R\right)
\chi_{c'}^{\left(  J\right)  }\left(  K_{n'I'},R\right)  ,
\label{3cceq}%
\end{align}
where $\mu$ is the $N$-$^6$Li reduced mass. Here, $c$ represents the set of $n$, $I$, and $L$. The relative wave number $K_{nI}$ is defined by%
\begin{equation}
K_{nI}=\frac{\sqrt{2\mu E_{nI}}}{\hbar}
\end{equation}
with
\begin{equation}
E_{nI}=E_{\rm tot}-\epsilon_{nI},
\end{equation}
where $E_{\rm tot}$ is the total energy.

The coupling potential is given in the single-folding model by
\begin{align}
U^{(J)}_{cc'}\left(  R\right) = &
\left.\left\langle \left[\Phi_{nI}\left(  \xi\right)  \otimes i^{L%
}Y_{L}(  \boldsymbol{\hat{R}} )  \right]  _{JM_J}\right\vert A g_{NN} \right. \nonumber \\
& \times \left\vert \left[\Phi_{n'I'}\left(  \xi\right)
\otimes i^{L'}Y_{L'}(  \boldsymbol{\hat{R}} )  \right]  _{JM_J}
\right\rangle _{\xi,\boldsymbol{\hat{R}}},
\label{3ucc}%
\end{align}
where $Y_L$ is the spherical harmonics. $A$ is the number of nucleons contained in $^6$Li, i.e., $A=6$, and $g_{NN}$ is a nucleon-nucleon ($NN$) $g$-matrix interaction. $g_{NN}$ is a function of the relative coordinate between the incident nucleon and a nucleon in $^6$Li, as well as the nuclear density at which the $NN$ collision occurs. The coupling potential can be written by the transition densities of $^6$Li. For details, readers are referred to Ref.~\cite{Hag22}.

The CC equations \eqref{3cceq} are solved under the following boundary conditions:
\begin{equation}
\chi_{c}^{(J)}
\to
U^{(-)}_{c} (K_{nI} R)\delta_{cc_0}
-\sqrt{\dfrac{K_{0}}{K_{nI}}}\;
S_{cc_0}
U^{(+)}_{c} (K_{nI} R)
\label{bc1}
\end{equation}
for $E_{nI} > 0$, and
\begin{equation}
\chi_c^{(J)}
\to
-S_{cc_0}
W_{-\eta_{nI},L+1/2} (-2i K_{nI} R)
\label{bc2}
\end{equation}
for $E_{nI} \le 0$; $c_0$ represents the incident channel and $K_0$ is the corresponding $N$-$^6$Li relative wave number.
$U^{(-)}_{c}$ ($U^{(+)}_{c}$) is the incoming (outgoing) Coulomb wave function and
$W_{-\eta_{nI},L+1/2}$ is the Whittaker function with the Sommerfeld parameter $\eta_{nI}$.
Channels having  $E_{nI} > 0$ and $E_{nI} \le 0$ are called
open and closed channels, respectively.
$S_{cc_0}$ for open channels are scattering matrix elements,
with which physics observables are calculated in a standard manner.
On the other hand, $S_{cc_0}$ for closed channels
are not related to observables,
at least directly. Nevertheless, the closed channels can affect the observables through mainly continuum-continuum
couplings~\cite{Aus96,OY16}.

\begin{figure}[!htpb]
\includegraphics[width=0.95\hsize]{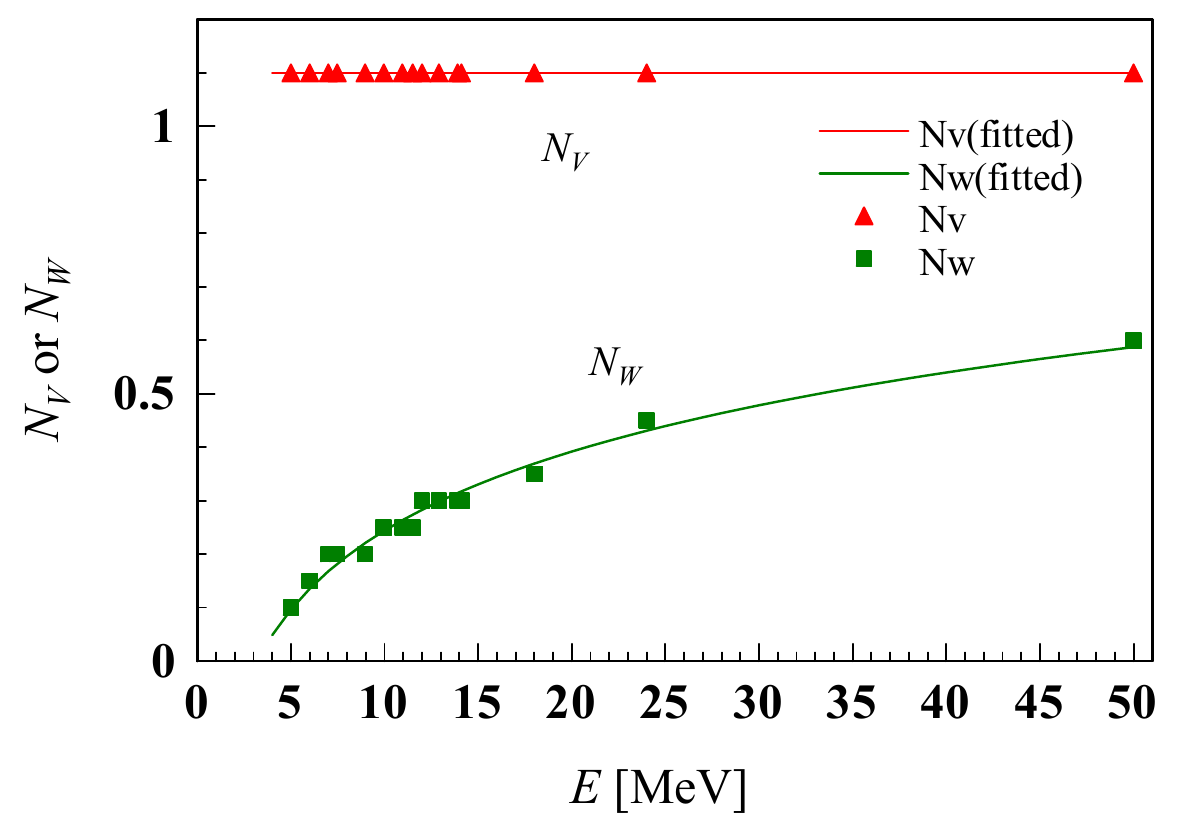}
\caption{Renormalization factors of the JLM interaction for the real ($N_V$) and imaginary ($N_W$) parts. The horizontal axis is the incident energy. The values obtained with the functional fitting are shown by the lines.}
\label{fig:nvnw}
\end{figure}
\section{Numerical inputs}
\label{sec:num}

\begin{figure*}[!htpb]
\includegraphics[width=0.7\hsize]{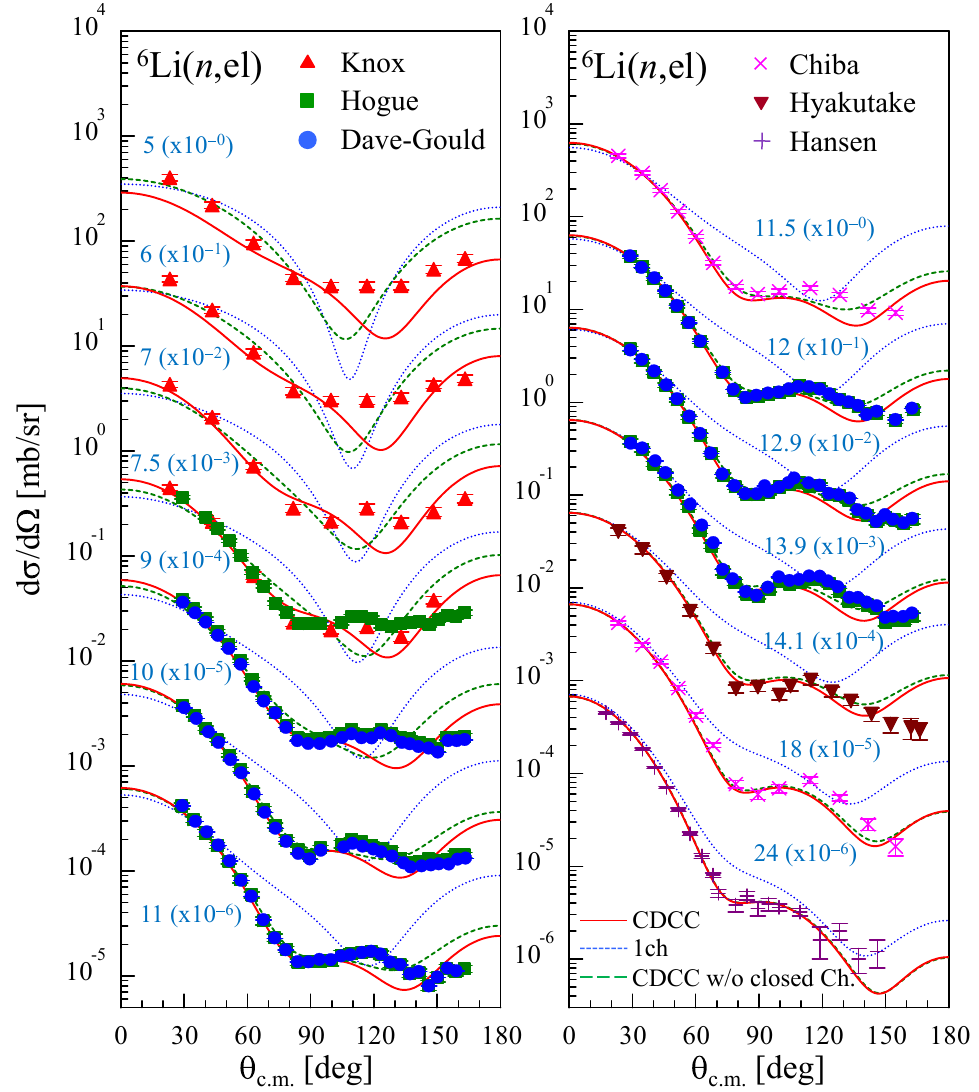}
\caption{
Angular distributions of neutron elastic scattering compared with experimental data. The numbers shown near the plot show the incident energies in MeV. At each energy, the red solid, green dashed, and blue dotted lines show the results of four-body CDCC, that without the closed channels, and the result of the single channel calculation, respectively. For CDCC without the closed channels, $N_{\rm st}=10$, 14, 24, 35, 55, 69, 77, 88, 91, 100, 123, 128, 179, 227 at 5, 6, 7, 7.5, 9, 10, 11, 11.5, 12, 12.9, 13.9, 14.1, 18, and 24~MeV, respectively.
Experimental data are taken from Refs.~\cite{Kno79,Hog79,DG83,Chi98,Hya74,Han88}.
}
\label{fig:nelas}
\end{figure*}
We use an $\alpha+p+n$ three-body model to obtain the transition densities of $^6$Li. The three-body Hamiltonian $h$ and the basis functions are the same as in Ref.~\cite{Wat15} except for the range $\nu$ and depth $V_3$ of the effective three-body interaction; see Eq.~(27) of Ref.~\cite{Wat15}. We use $\nu=0.036$~fm$^{-2}$, whereas $V_3$ is determined to reproduce the ground-state energy for the $1^+$ state and the low-lying resonance energies for the $2^+$ and $3^+$ states.

\begin{figure}[!htpb]
\includegraphics[width=0.95\hsize]{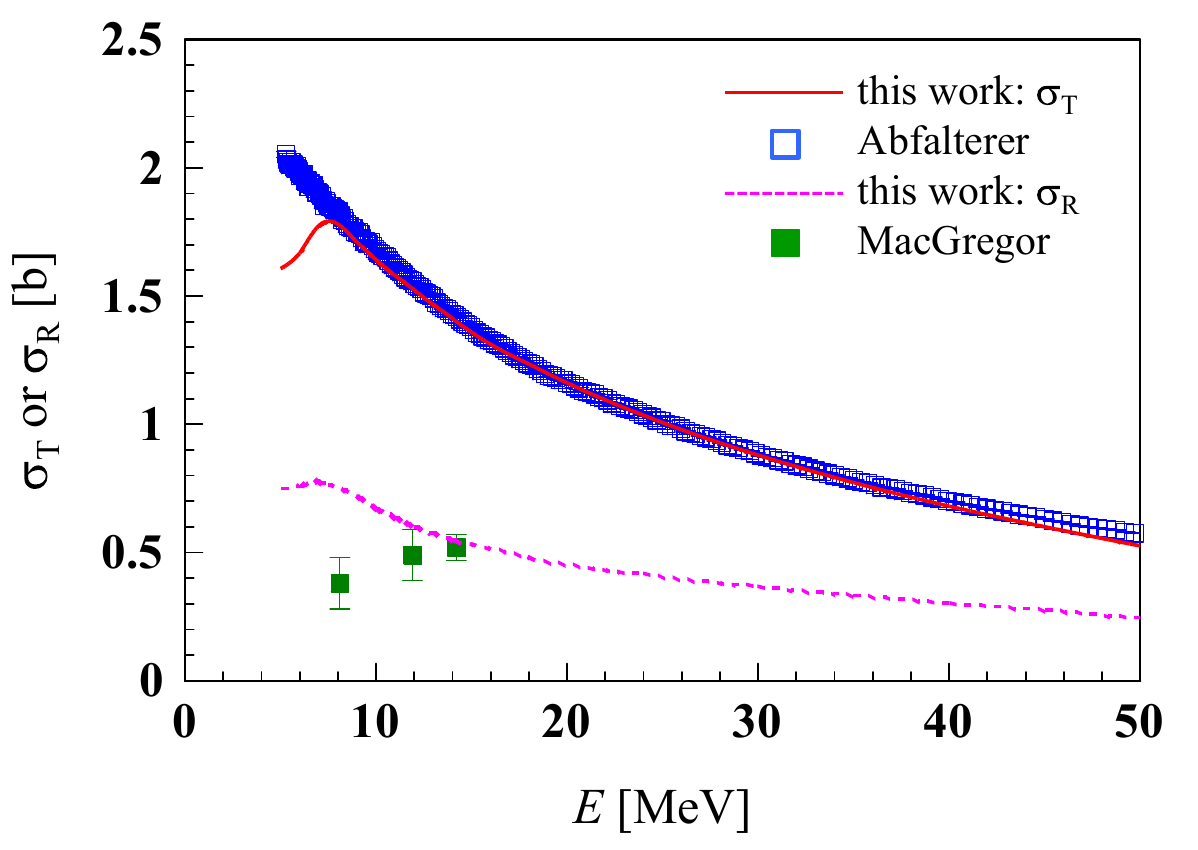}
\caption{Neutron total and total-reaction cross sections as a function of incident energy. Experimental data are taken from Refs.~\cite{Abf01,Mac63}}
\label{fig:ntot}
\end{figure}
After diagonalization of $h$ with the basis functions, we got 2456 eigenstates of $^6$Li, out of which we adopted $\Phi_{nIm_{I}}$ having $\epsilon_{nI} \le 20$~MeV in the CDCC calculation. Consequently, we have implemented 79, 83, and 92 states for the $1^+$, $2^+$, and $3^+$ states of $^6$Li, respectively; the total number of states $N_{\rm st}$ is 254.

The boundary conditions of Eqs.~(\ref{bc1}) and (\ref{bc2}) are imposed at $R=20$~fm. The maximum orbital angular momentum, $L_{\rm max}$, is chosen depending on the incident energy $E$; we set $L_{\rm max}$ to 5 (23) for the 4~MeV (72~MeV) scattering. These calculation settings give the convergence of the results shown below.

\begin{figure*}[!htpb]
\includegraphics[width=0.7\hsize]{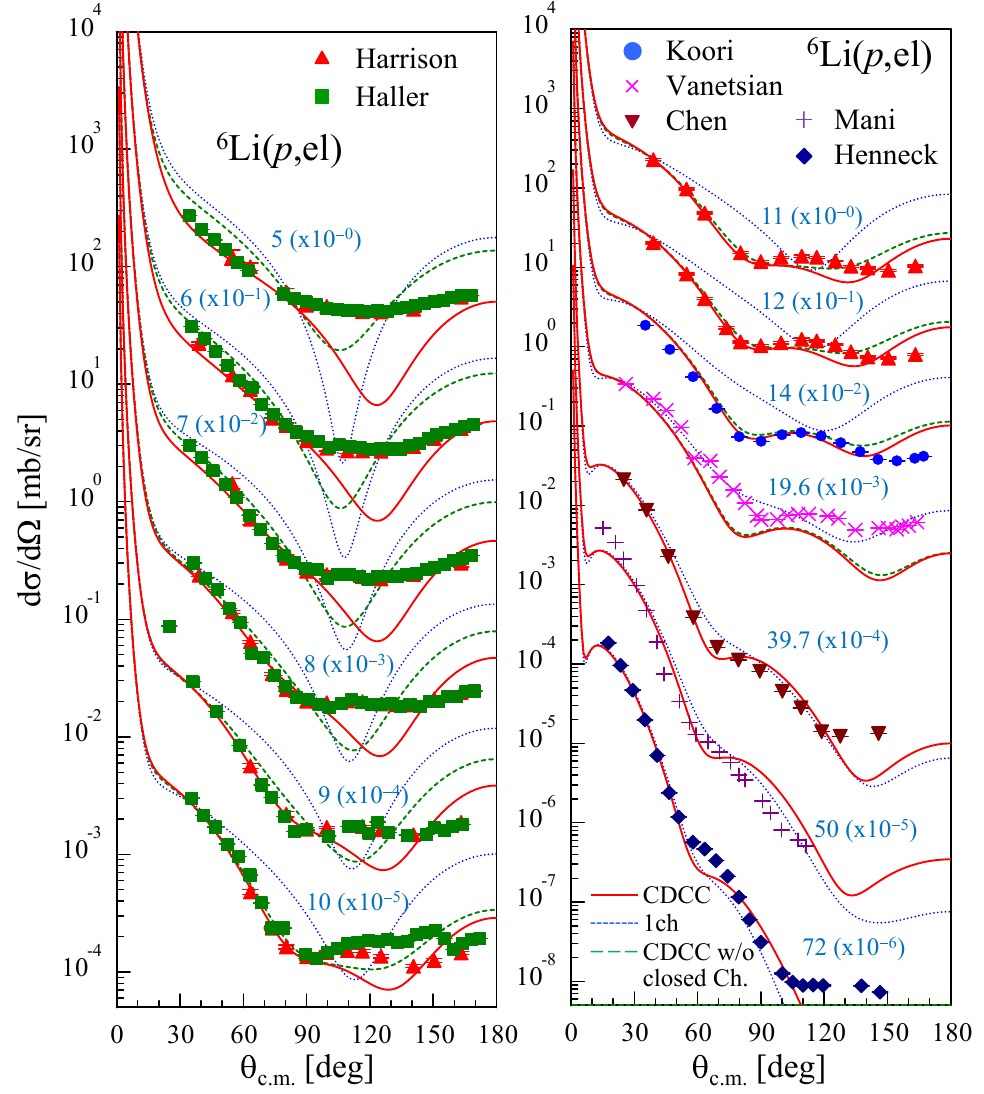}
\caption{Same as Fig.~\ref{fig:nelas} but for proton elastic scattering. For CDCC without the closed channels, $N_{\rm st}=10$, 14, 24, 42, 35, 69, 77, 91, 126, 192 at 5, 6, 7, 8, 9, 10, 11, 12, 14, and 19.6~MeV, respectively; at 39.7~MeV and above, all channels are open.
Experimental data are taken from Refs.~\cite{HW63,Hal89,Koo89,Van60,SH60,Man67,Hen94}.}
\label{fig:pelas}
\end{figure*}
As for the $g$-matrix interaction, we employ the parametrization of JLM~\cite{Jeu77}; the ranges of the Gaussian for both real and imaginary parts are set to 1.2~fm, following the standard choice. On the other hand, the renormalization factors $N_V$ and $N_W$ for the real and imaginary parts, respectively, are treated as free parameters. Note that although $N_V=1.0$ and $N_W=0.8$ are suggested as standard values, these are based on the optical model calculation. Because we adopt the CC formalism, some changes of the parameters are expected naturally.

\section{Results and Discussion}
\label{sec:res}
\begin{figure}[!htpb]
\includegraphics[width=0.95\hsize]{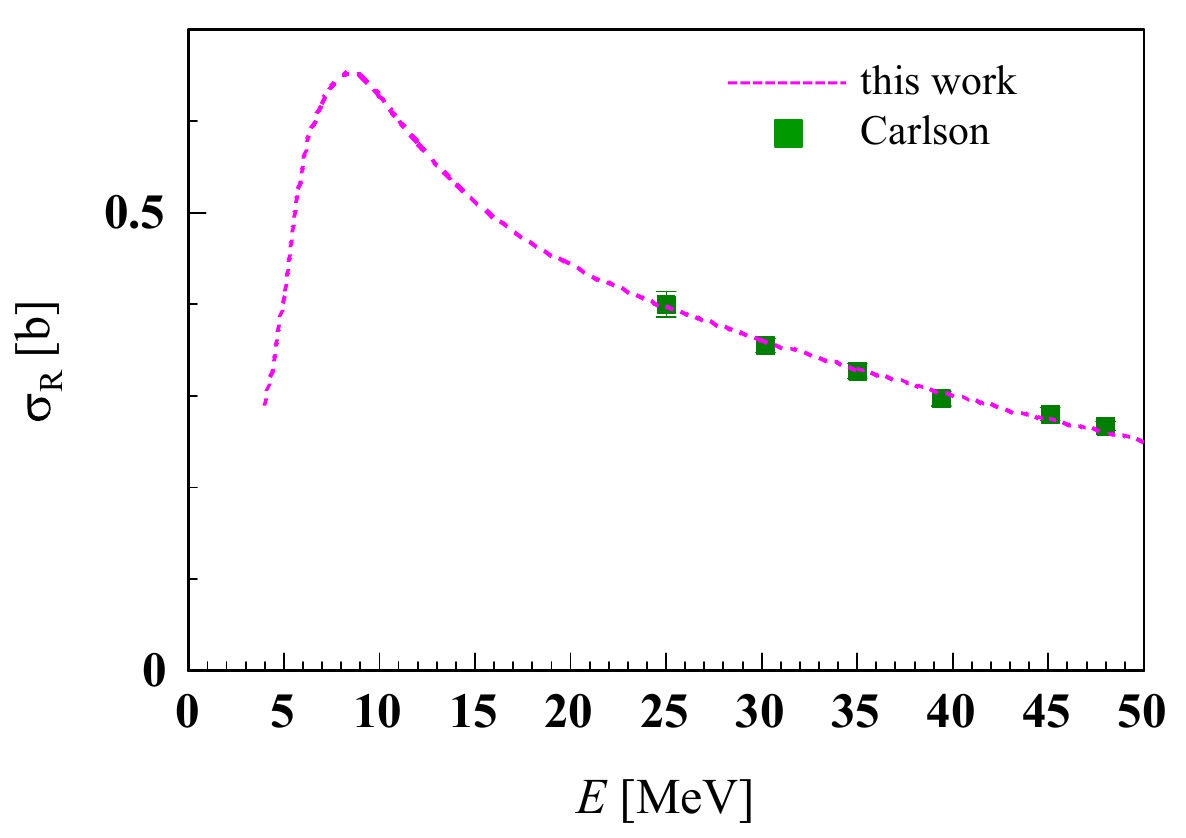}
\caption{Proton total reaction cross sections as a function of incident energy. Experimental data are taken from Ref.~\cite{Car85}.}
\label{fig:preac}
\end{figure}
To determine $N_V$ and $N_W$, we have calculated the $n$-$^6$Li elastic scattering cross sections at 14 energies between 5~MeV and 24~MeV with $N_V=1.0,$ 1.1, and 1.2; for each value of $N_V$, we changed $N_W$ from 0.0 to 1.0 with an increment of 0.05. Thus, we tried 63 sets of $(N_V, N_W)$ at each energy. Note that the four-body CDCC calculation is rather computationally demanding, so the $\chi^2$ fitting of $(N_V,N_W)$ is practically difficult. The {\lq\lq}best'' values of $N_v$ and $N_W$ thus determined are shown in Fig.~\ref{fig:nvnw}. In the figure, the results for $p$-$^6$Li at 50~MeV, which is needed for determining the energy dependence of the parameters above 24~MeV, are also plotted. One sees that the constant value of 1.1 is the optimal renormalization factor for the real part $N_V$. On the other hand, $N_W$ is a smooth function of the incident energy $E$ and can be fitted by
\begin{equation}
N_W =0.213\ln{E}-0.247.
\label{eq:nw}
\end{equation}
Note that $N_W$ goes to 0.84 at 160~MeV, which is the upper limit of the incident energy of the JLM $g$-matrix interaction. 
It was found that, at 24~MeV, several sets of $(N_V, N_W)$ reproduce the $n$-$^6$Li elastic scattering cross section with almost the same quality. However, only one of these sets also reproduces $\sigma^{\rm tot}_{n{}^{6}{\rm Li}}$. We therefore identify $N_V=1.1$ and $N_W$ given by Eq.~\eqref{eq:nw} as the optimal parameter set and use these values in the following analysis.

We show in Fig.~\ref{fig:nelas} the results of the $n$-$^6$Li elastic scattering cross sections at 14 energies, together with the experimental data. The red solid lines are the four-body CDCC results, which satisfactorily reproduce the experimental data, although slight undershooting is found around 130$^\circ$ below 14~MeV. The blue dotted lines are the results without breakup channels, which obviously fail to reproduce the data. This clearly shows the importance of the breakup channels in the description of the elastic scattering. The green dashed lines represent the four-body CDCC results without the closed channels. As expected, the role of the closed channels is significant at low energies, whereas it tends to less important as $E$ increases.

In the preceding study with three-body CDCC~\cite{Mat11}, the closed channels were omitted and $(N_V,N_W)=(1.0,0.1)$ were adopted at all energies; the data down to 7.5~MeV were analyzed. The agreement with experimental data were quite well, but slight overshooting at forward angles was found. In a naive expectation, because the breakup channels considered in the present study cover wider model space than in Ref.~\cite{Mat11}, the overall factor for the absorption, $N_W$, would become smaller. However, this is not the case as shown in Fig.~\ref{fig:nvnw}. At low energies, as discussed above, the effect coming from the closed channels cannot be neglected. Thus, it can be interpreted that the role of the closed-channels was effectively taken into account in $N_W$ in the previous study~\cite{Mat11}. Because the closed channels have no outgoing fluxes, they do not directly contribute to the absorption. Therefore, a naive comparison between the values of $N_W$ in the present study and in Ref.~\cite{Mat11} will not make sense. At 24~MeV, the effect of the closed channels becomes negligible. Although $(N_V,N_W)=(1.1,0.43)$ are the optimal values for the present model, four-body CDCC using $(N_V,N_W)=(1.0,0.1)$ is found to be rather close to that in Ref.~\cite{Mat11}. However, this choice of $N_V$ and $N_W$ was found not to explain the $n$-$^6$Li total cross section $\sigma^{\rm tot}_{n{}^{6}{\rm Li}}$ above 25~MeV. As shown below, with $(N_V,N_W)$ in Fig.~\ref{fig:nvnw}, we can reproduce $\sigma^{\rm tot}_{n{}^{6}{\rm Li}}$ up to 50~MeV.

Figure~\ref{fig:ntot} displays the comparison between the calculated $\sigma^{\rm tot}_{n{}^{6}{\rm Li}}$ (crosses) and the experimental data (open squares); the result for the neutron total reaction cross sections $\sigma^{\rm R}_{n{}^{6}{\rm Li}}$ (plus signs) and the experimental data (closed squares) are also shown.
One sees that the calculation reproduces well $\sigma^{\rm tot}_{n{}^{6}{\rm Li}}$ above 7~MeV. The data for $\sigma^{\rm R}_{n{}^{6}{\rm Li}}$ at 8~MeV is severely overshot, which is also the case with phenomenological approaches. Based on the results in Fig.~\ref{fig:ntot}, we conclude that the present four-body CDCC model is applicable down to 7~MeV. In fact, as shown in Fig.~\ref{fig:nelas}, at 5 and 6 MeV, the elastic scattering cross sections cannot be reproduced even at forward angles. At energies below 7~MeV, the semi-microscopic folding model will not be applicable and more sophisticated many-body approaches will be necessary. It is also found that above 50~MeV, four-body CDCC tends to undershoot $\sigma^{\rm tot}_{n{}^{6}{\rm Li}}$. Although the present model can describe the $p$-$^6$Li elastic scattering cross section at 72~MeV (see Fig.\ref{fig:pelas}), it can safely be applied to the scattering up to 50~MeV.

The results for the $p$-$^6$Li elastic scattering are shown in Fig.~\ref{fig:pelas} and those for the total reaction cross sections $\sigma^{\rm R}_{p{}^{6}{\rm Li}}$ are in Fig.~\ref{fig:preac}.
It should be noted that except at 50~MeV, the $p$-$^6$Li observables have not been used in the determination of $(N_V,N_W)$ in Fig~\ref{fig:nvnw}. The quality of the agreement with data is similar to that for neutron scattering, and also to the preceding phenomenological study~\cite{Ye08}. The rather severe undershooting around 120$^\circ$ at lower energies may be attributed to the compound elastic component, which needs further investigation. We remark that the elastic scattering cross section at 72~MeV is well reproduced. This indicates that the extrapolation of $(N_V,N_W)$ shown in Fig.~\ref{fig:nvnw} works well up to this energy. However, as mentioned, there is no guarantee that the present model can describe the total reaction cross section at 72~MeV.

\section{summary}
\label{sec:sum}
We have constructed a semi-microscopic model for nucleon-$^6$Li scattering for the energies from 7~MeV to 50~MeV. The model is based on the four-body continuum-discretized coupled-channels method with an $\alpha+p+n$ three-body model. The nucleon-nucleon effective $g$-matrix interaction by Jeukenne, Lejeune, and Mahaux (JLM) was employed. We have determined the parameters of the JLM interaction to reproduce the neutron elastic scattering cross sections at 14 energies up to 24~MeV, and the proton cross section at 50~MeV. The parameters thus determined are found to reproduce the proton elastic scattering cross sections at eleven energies below 50~MeV as well as that at 72~MeV. The total and total-reaction cross sections are also well reproduced by the present four-body CDCC calculation between 7~MeV and 50~MeV. At the energies outside this range, the present model undershoots the neutron total cross section, which indicates the limitation of the present semi-microscopic approach with the JLM interaction.

At energies below 7~MeV, a more sophisticated many-body approach will be needed, whereas above 50~MeV, one may use the MST-based microscopic approach. For example, a microscopic folding model with the Melbourne $g$-matrix interaction can be used; it has been successful in describing various nucleon-nucleus scatterings between 65~MeV and 200~MeV with no free adjustable parameters.

Because we use four-body CDCC, the scattering matrices to discretized breakup channels have already been evaluated. To generate physics observables that are continuous regarding the breakup energy, however, we need a smoothing technique such as the complex-scaling smoothing. Another important ingredient is the channel selection. In the present calculation, the discretized continuum states of $^6$Li contain both $\alpha+d$ and $\alpha+p+n$ components, which must be disentangled to make a comparison with data. Very recently, a channel selection combining four-body CDCC and the method of the complex-scaled solution of the Lippmann-Schwinger equation (CSLS) has successfully been applied to the breakup of $^9$C~\cite{Oga25}. With these methods, we will report results of breakup processes of $^6$Li in a forthcoming paper.

Considering the importance in the nuclear data science, neutron reactions with $^7$Li will be more emphasized. A similar model to the present model can be constructed, once an $\alpha+p+n+n$ four-body model description becomes available. A collaboration with E.~Hiyama (Tohoku University/RIKEN) is ongoing and some results will appear in the near future.

\begin{acknowledgments}
The authors thank K.~Sekiguchi, E.~Hiyama, H.~Otsu, O.~Iwamoto, N.~Iwamoto, T.~Fukahori, S.~Shimoura, Y.~Watanabe, S.~Watanabe, and T.~ Matsumoto for fruitful discussions. This work was supported in part by JST ERATO Grant No. JPMJER2304, Japan, and by Grants-in-Aid of the Japan Society for the Promotion of Science (Grants No. JP25K07302). The computation was carried out with the computer facilities at the Research Center
for Nuclear Physics, the University of Osaka.
\end{acknowledgments}

\appendix

\section{Qualification of the fitting of $N_V$ and $N_W$}

\begin{figure}[!htpb]
\includegraphics[width=0.95\hsize]{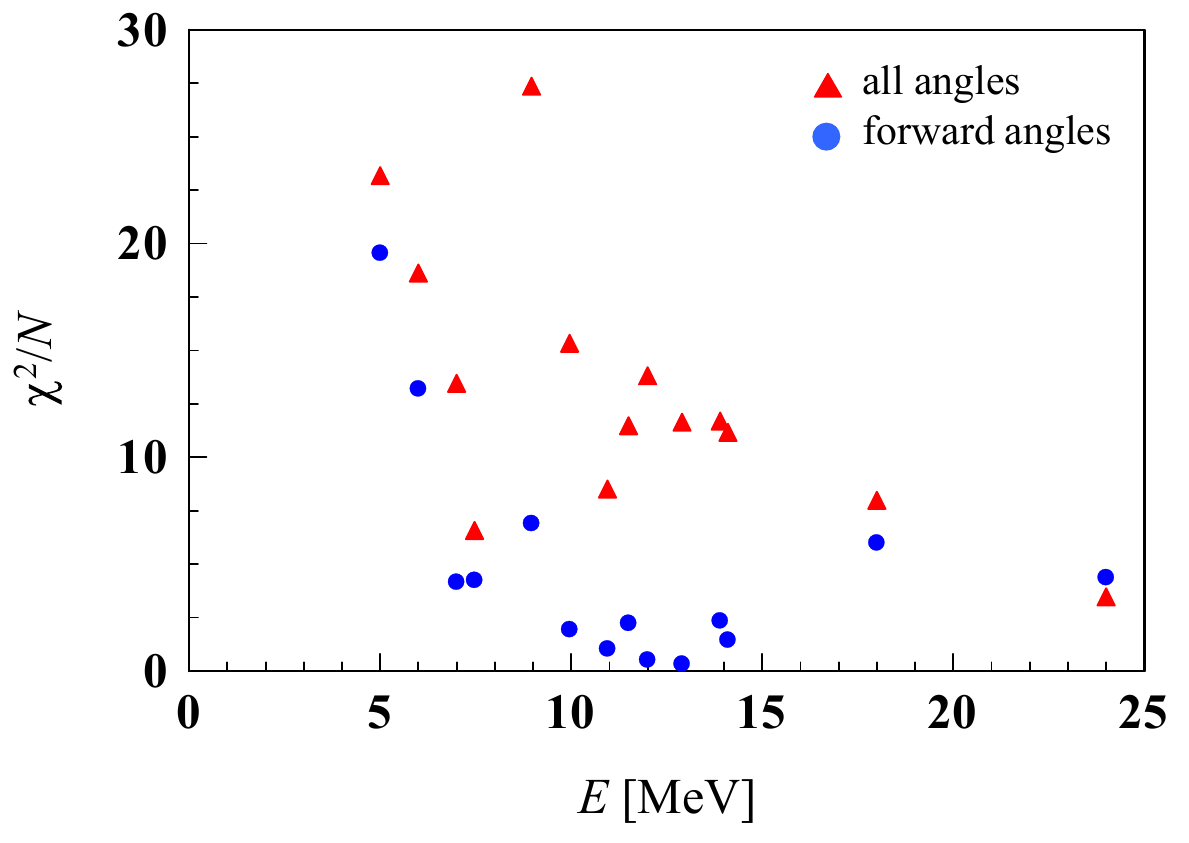}
\caption{$\chi^2/N$ for $n$-$^6$Li elastic scattering as a function of incident energy. Triangles and circles are the results using the experimental data at all angles and those up to $90^\circ$, respectively}
\label{fig:chi2}
\end{figure}
As mentioned above, we have not determined $(N_V,N_W)$ through a formal $\chi^2/N$ minimization. Therefore, statistical uncertainties in the fitted parameters cannot be evaluated within the present analysis. Nevertheless, it is informative to quantify the agreement between the experimental data and the four-body CDCC calculations using the {\lq\lq}fitted'' values of $(N_V,N_W)$ shown in Fig.~\ref{fig:nvnw}. The values of $\chi^2/N$ presented below should thus be regarded as a measure of the quality of the agreement rather than as the result of a statistical parameter estimation.

We show in Fig.~\ref{fig:chi2}
\begin{equation}
    \chi^2/N = \sum_{i=1}^N \frac{[(d\sigma/d\Omega_i)_{\rm theo}-(d\sigma/d\Omega_i)_{\rm exp}]^2}{[(\Delta d\sigma/d\Omega_i)_{\rm exp}]^2}
    \label{eq:chi2}
\end{equation}
for the neutron elastic scattering cross sections in Fig.~\ref{fig:nelas}; $(d\sigma/d\Omega_i)_{\rm theo}$ and $(d\sigma/d\Omega_i)_{\rm exp}$ denote the theoretical and experimental $n$-$^6$Li differential cross sections, respectively, at the measured angle $\theta_i$, and $(\Delta d\sigma/d\Omega_i)_{\rm exp}$ denotes the statistical uncertainty in $(d\sigma/d\Omega_i)_{\rm exp}$. The triangles and circles represent the values of $\chi^2/N$ obtained using data at all angles and only at angles up to $90^\circ$, respectively. The horizontal axis is the neutron incident energy $E$.

As a general trend, $\chi^2/N$ decreases with increasing $E$, probably reflecting the energy dependence of the reliability of the MST. Restricting the analysis to the forward-angle data reduces $\chi^2/N$ except at 24~MeV. This reduction is particularly significant at energies between 9 and 14.1~MeV. It is well known that reproducing elastic scattering cross sections at very backward angles is nontrivial. Therefore, the difference between the triangles and circles can be understood naturally. The somewhat larger values of $\chi^2/N$ at 18 and 24~MeV may indicate the need for a more careful statistical analysis. Such an analysis could employ not only the $\chi^2/N$ measure defined in Eq.~\eqref{eq:chi2} but also Bayesian statistical methods, and is left for future work.


\bibliographystyle{apsrev4-2}
\bibliography{ko}

@article{Mat11,
  title = {Systematic description of the $^{6}\mathrm{Li}$($n,{n}^{\ensuremath{'}}){}^{6}$Li${}^{*}\ensuremath{\rightarrow}d+\ensuremath{\alpha}$ reactions with the microscopic coupled-channels method},
  author = {Matsumoto, T. and Ichinkhorloo, D. and Hirabayashi, Y. and Kat\ifmmode \bar{o}\else \={o}\fi{}, K. and Chiba, S.},
  journal = {Phys. Rev. C},
  volume = {83},
  issue = {6},
  pages = {064611},
  numpages = {6},
  year = {2011},
  month = {Jun},
  publisher = {American Physical Society},
  doi = {10.1103/PhysRevC.83.064611},
  url = {https://link.aps.org/doi/10.1103/PhysRevC.83.064611}
}

@article{Kam86,
  author  = {M. Kamimura and M. Yahiro and Y. Iseri and Y. Sakuragi and H. Kameyama and M. Kawai},
  journal = {Prog. Theo. Phys. Suppl.},
  volume  = {89},
  pages   = {1},
  year    = {1986}
}

@article{Aus87,
  author  = {N. Austern and Y. Iseri and M. Kamimura and M. Kawai and G. Rawitscher and M. Yahiro},
  journal = {Phys. Rep.},
  volume  = {154},
  pages   = {125},
  year    = {1987}
}

@article{Yah12,
  author  = {M. Yahiro and K. Ogata and T. Matsumoto and K. Minomo},
  journal = {Prog. Theo. Exp. Phys.},
  volume  = {2012},
  pages   = {01A206},
  year    = {2012}
}

@article{Fol45,
  author  = {L. L. Foldy},
  journal = {Phys. Rev.},
  volume  = {67},
  pages   = {107},
  year    = {1945}
}

@article{Wat53,
  author  = {K. M. Watson},
  journal = {Phys. Rev.},
  volume  = {89},
  pages   = {115},
  year    = {1953}
}

@article{Ker59,
  author       = {Kerman, A. K. and McManus, H. and Thaler, R. M.},
  title        = {The scattering of fast nucleons from nuclei},
  journal      = {Annals of Physics},
  volume       = {8},
  pages        = {551--635},
  year         = {1959},
  publisher    = {Elsevier},
  note         = {Foundational multiple scattering / optical potential for nucleon--nucleus scattering; basis of KMT formalism} 
}

@incollection{Amo00,
  title        = {Nucleon-Nucleus Scattering: A Microscopic Nonrelativistic Approach},
  author       = {Amos, K. and Dortmans, P. J. and von Geramb, H. V. and Karataglidis, S. and Raynal, J.},
  booktitle    = {Advances in Nuclear Physics},
  volume       = {25},
  pages        = {276--536},
  publisher    = {Springer},
  year         = {2000},
  doi          = {10.1007/0-306-47101-9_3}
}

@article{OY16,
  author  = {K. Ogata and K. Yoshida},
  journal = {Phys. Rev. C},
  volume  = {94},
  pages   = {051603(R)},
  year    = {2016}
}

@article{Toy13,
  author  = {M. Toyokawa and K. Minomo and M. Yahiro},
  journal = {Phys. Rev. C},
  volume  = {88},
  pages   = {054602},
  year    = {2013}
}

@article{Toy15,
  author  = {M. Toyokawa and M. Yahiro and T. Matsumoto and K. Minomo and K. Ogata and M. Kohno},
  journal = {Phys. Rev. C},
  volume  = {92},
  pages   = {024618},
  year    = {2015}
}

@article{Jeu77,
  author  = {J.-P. Jeukenne and A. Lejeune and C. Mahaux},
  journal = {Phys. Rev. C},
  volume  = {16},
  pages   = {80},
  year    = {1977}
}

@article{Aus96,
  author  = {N. Austern and M. Kawai and M. Yahiro},
  journal = {Phys. Rev. C},
  volume  = {53},
  pages   = {314},
  year    = {1996}
}

@article{Hiy03,
  author  = {E. Hiyama and Y. Kino and M. Kamimura},
  journal = {Prog. Part. Nucl. Phys.},
  volume  = {51},
  pages   = {223},
  year    = {2003}
}

@article{Wat15,
  author  = {S. Watanabe and T. Matsumoto and K. Ogata and M. Yahiro},
  journal = {Phys. Rev. C},
  volume  = {92},
  pages   = {044611},
  year    = {2015}
}

@article{Mat04,
  author = {Matsumoto, T. and Hiyama, E. and Ogata, K. and Iseri, Y. and Kamimura, M. and Chiba, S. and Yahiro, M.},
  journal = {Phys. Rev. C},
  volume = {70},
  pages = {061601},
  year = {2004}
}

@article{Hag22,
title = {Coupled-channels calculations for nuclear reactions: From exotic nuclei to superheavy elements},
journal = {Progress in Particle and Nuclear Physics},
volume = {125},
pages = {103951},
year = {2022},
issn = {0146-6410},
doi = {https://doi.org/10.1016/j.ppnp.2022.103951},
url = {https://www.sciencedirect.com/science/article/pii/S0146641022000126},
author = {K. Hagino and K. Ogata and A.M. Moro}
}

@article{Kno79,
  author  = {H. D. Knox and R. M. White and R. O. Lane},
  journal = {Nucl. Sci. Eng.},
  volume  = {69},
  pages   = {223},
  year    = {1979},
  note    = {; EXFOR-10710 data file entry}
}

@article{Hog79,
  author  = {H. H. Hogue and P. L. von Behren and D. W. Glasgow and S. G. Glendinning and P. W. Lisowski and C. E. Nelson and F. O. Purser and W. Tornow and C. R. Gould and L. W. Seagondollar},
  journal = {Nucl. Sci. Eng.},
  volume  = {69},
  pages   = {22},
  year    = {1979},
  note    = {; EXFOR-10707 data file entry}
}

@article{DG83,
  title = {Optical model analysis of scattering of 7- to 15-MeV neutrons from $1\ensuremath{-}p$ shell nuclei},
  author = {Dave, J. H. and Gould, C. R.},
  journal = {Phys. Rev. C},
  volume = {28},
  issue = {6},
  pages = {2212--2221},
  numpages = {0},
  year = {1983},
  month = {Dec},
  publisher = {American Physical Society},
  doi = {10.1103/PhysRevC.28.2212},
  url = {https://link.aps.org/doi/10.1103/PhysRevC.28.2212}
}

@article{Chi98,
  author  = {S. Chiba and K. Togasaki and M. Ibaraki and M. Baba and S. Matsuyama and N. Hirakawa and K. Shibata and O. Iwamoto and A. J. Koning and G. M. Hale and M. B. Chadwick},
  journal = {Phys. Rev. C},
  volume  = {58},
  pages   = {2205},
  year    = {1998},
  note    = {}
}

@article{Hya74,
  author  = {M. Hyakutake and M. Sonoda and A. Katase and Y. Wakuta and M. Matoba and H. Tawara and I. Fujita},
  journal = {J. Nucl. Sci. Technol.},
  volume  = {11},
  pages   = {407},
  year    = {1974},
  note    = {; EXFOR-20268 data file entry}
}

@article{Han88,
  author  = {L. F. Hansen and J. Rapaport and X. Wang and F. A. Barrios and F. Petrovich and A. W. Carpenter and M. J. Threapleton},
  journal = {Phys. Rev. C},
  volume  = {38},
  pages   = {525},
  year    = {1988},
  note    = {; EXFOR-13161 data file entry}
}

@article{Abf01,
  author  = {W. P. Abfalterer and F. B. Bateman and F. S. Dietrich and R. W. Finlay and R. C. Haight and G. L. Morgan},
  journal = {Phys. Rev. C},
  volume  = {63},
  pages   = {044608},
  year    = {2001},
  note    = {; EXFOR-13753 data file entry}
}

@article{Mac63,
  author  = {M. H. Mac Gregor and R. Booth and W. P. Ball},
  journal = {Phys. Rev.},
  volume  = {130},
  pages   = {1471},
  year    = {1963},
  note    = {; EXFOR-11120 data file entry}
}

@article{HW63,
  author  = {W. D. Harrison and A. B. Whitehead},
  journal = {Phys. Rev.},
  volume  = {132},
  pages   = {2607},
  year    = {1963},
  note    = {; EXFOR-C1003 data file entry}
}

@article{Hal89,
  author  = {M. Haller and W. Kretschmer and A. Rauscher and R. Schmitt and W. Schuster},
  journal = {Nucl. Phys. A},
  volume  = {496},
  pages   = {189},
  year    = {1989},
  note    = {; EXFOR-F0063 data file entry}
}

@article{Koo89,
  author  = {N. Koori and I. Kumabe and M. Hyakutake and K. Orito and K. Akagi and A. Iida and Y. Watanabe and K. Sagara and H. Nakamura and K. Maeda and T. Nakashima and M. Kamimura and Y. Sakuragi},
  journal = {JEARI-M},
  volume  = {89},
  pages   = {167},
  year    = {1989},
  note    = {}
}

@article{Van60,
  author  = {R. A. Vanetsian and A. P. Klyucharev and E. D. Fedchenko},
  journal = {Sov. At. Energ.},
  volume  = {6},
  pages   = {490},
  year    = {1960},
  note    = {}
}

@InProceedings{SH60,
author = {S. Chen and N. M. Hintz},
booktitle  = {inInternational Conference on Nuclear Forces and the Few Nucleon Problem},
editor = {T. C. Griffith and E. A. Power},
pages = {683},
publisher = {Pergamon Press},
address = {New York},
year={1960}
}

@article{Man67,
  author  = {G. S. Mani and A. D. B. Dix and D. T. Jones and M. Richardson},
  journal = {Rutherford Laboratory Report},
  volume  = {No. RHEL/R-136},
  pages   = {49},
  year    = {1967},
  note    = {(unpublished)}
}

@article{Hen94,
title = {Depolarization in proton-6Li elastic scattering},
journal = {Nuclear Physics A},
volume = {571},
number = {3},
pages = {541-554},
year = {1994},
issn = {0375-9474},
doi = {https://doi.org/10.1016/0375-9474(94)90224-0},
url = {https://www.sciencedirect.com/science/article/pii/0375947494902240},
author = {R. Henneck and G. Masson and P.D. Eversheim and R. Gebel and F. Hinterberger and U. Lahr and H.W. Schmitt and J. Schleef and B.v. Przewoski}
}

@article{Car85,
  author  = {R. F. Carlson and A. J. Cox and T. N. Nasr and M. S. De Jong and D. L. Ginther and D. K. Hasell and A. M. Sourkes and W. T. H. van Oers and D. J. Margaziotis},
  journal = {Nucl. Phys. A},
  volume  = {445},
  pages   = {57},
  year    = {1985},
  note    = {; EXFOR-C0215 data file entry}
}

@InProceedings{Mas04,
author = {H. Matsui},
booktitle  = {Proceedings of the 23rd Symposium on Fusion Technology},
editor = {},
pages = {},
publisher = {},
address = {},
year={},
note={Venice, Italy, 20–24 September 2004}
}

@article{Fen24,
title = {FENDL: A library for fusion research and applications},
journal = {Nuclear Data Sheets},
volume = {193},
pages = {1-78},
year = {2024},
note = {Special Issue on Nuclear Reaction Data},
issn = {0090-3752},
doi = {https://doi.org/10.1016/j.nds.2024.01.001},
url = {https://www.sciencedirect.com/science/article/pii/S0090375224000012},
author = {G. Schnabel and D.L. Aldama and T. Bohm and U. Fischer and S. Kunieda and A. Trkov and C. Konno and R. Capote and A.J. Koning and S. Breidokaite and T. Eade and M. Fabbri and D. Flammini and L. Isolan and I. Kodeli and M. Kostal and S. Kwon and D. Laghi and D. Leichtle and S. Nakayama and M. Ohta and L.W. Packer and Y. Qiu and S. Sato and M. Sawan and M. Schulc and G. Stankunas and M. Sumini and A. Valentine and R. Villari and A. Zohar}
}

@article{Gys24,
  title = {$\mathit{Ab initio}$ investigation of the $^{7}\mathrm{Li}(p,{e}^{+}{e}^{\ensuremath{-}})^{8}\mathrm{Be}$ process and the X17 boson},
  author = {Gysbers, P. and Navr\'atil, P. and Kravvaris, K. and Hupin, G. and Quaglioni, S.},
  journal = {Phys. Rev. C},
  volume = {110},
  issue = {1},
  pages = {015503},
  numpages = {25},
  year = {2024},
  month = {Jul},
  publisher = {American Physical Society},
  doi = {10.1103/PhysRevC.110.015503},
  url = {https://link.aps.org/doi/10.1103/PhysRevC.110.015503}
}

@article{Bar13,
  author       = {Baroni, S. and Navr{\'a}til, P. and Quaglioni, S.},
  title        = {Ab initio description of the exotic unbound $^{7}$He nucleus},
  journal      = {Physical Review Letters},
  volume       = {110},
  pages        = {022505},
  year         = {2013},
  doi          = {10.1103/PhysRevLett.110.022505}
}

@article{Nav16,
  author       = {Navr{\'a}til, P. and Quaglioni, S. and Hupin, G. and Romero-Redondo, C. and Calci, A.},
  title        = {Unified ab initio approaches to nuclear structure and reactions},
  journal      = {Physica Scripta},
  volume       = {91},
  number       = {5},
  pages        = {053002},
  year         = {2016},
  doi          = {10.1088/0031-8949/91/5/053002}
}

@article{Hup14,
  author       = {Hupin, G. and Quaglioni, S. and Navr{\'a}til, P.},
  title        = {Ab initio many-body calculations of nucleon--$^{4}$He scattering with three-nucleon forces},
  journal      = {Physical Review C},
  volume       = {90},
  pages        = {061601},
  year         = {2014},
  doi          = {10.1103/PhysRevC.90.061601}
}

@article{Bur19,
  author       = {Burrows, Matthew and Elster, Ch. and Weppner, S. P. and Launey, K. D. and Maris, P. and Nogga, A. and Popa, G.},
  title        = {Ab initio folding potentials for nucleon-nucleus scattering based on NCSM one-body densities},
  journal      = {Physical Review C},
  volume       = {99},
  pages        = {044603},
  year         = {2019},
  doi          = {10.1103/PhysRevC.99.044603},
  note         = {}
}

@article{Bur20,
  author       = {Burrows, Matthew and Baker, R. B. and Elster, Ch. and Weppner, S. P. and Launey, K. D. and Maris, P. and Popa, G.},
  title        = {Ab initio leading order effective potentials for elastic nucleon-nucleus scattering},
  journal      = {Physical Review C},
  volume       = {102},
  pages        = {034606},
  year         = {2020},
  doi          = {10.1103/PhysRevC.102.034606},
  note         = {}
}

@article{Vor24,
  author       = {Vorabbi, Matteo and Barbieri, Carlo and Somà, Vincenzo and Finelli, Paolo and Giusti, Carlotta},
  title        = {Microscopic optical potentials for medium-mass isotopes derived at the first order of Watson multiple-scattering theory},
  journal      = {Physical Review C},
  volume       = {109},
  number       = {3},
  pages        = {034613},
  year         = {2024},
  doi          = {10.1103/PhysRevC.109.034613},
  note         = {}
}

@article{Ye08,
  title = {Analysis of deuteron elastic scattering from $^{6,7}\mathrm{Li}$ using the continuum discretized coupled channels method},
  author = {Ye, Tao and Watanabe, Yukinobu and Ogata, Kazuyuki and Chiba, Satoshi},
  journal = {Phys. Rev. C},
  volume = {78},
  issue = {2},
  pages = {024611},
  numpages = {12},
  year = {2008},
  month = {Aug},
  publisher = {American Physical Society},
  doi = {10.1103/PhysRevC.78.024611},
  url = {https://link.aps.org/doi/10.1103/PhysRevC.78.024611}
}

@article{Oga25,
    author = {Ogawa, Shoya and Fukui, Tokuro and Singh, Jagjit and Ogata, Kazuyuki},
    title = {Determination of S18 from the 9C Breakup Reaction Within a Four-body Reaction Model},
    journal = {Progress of Theoretical and Experimental Physics},
    volume = {2026},
    number = {7},
    pages = {073D03},
    year = {2026},
    month = {07},
    issn = {2050-3911},
    doi = {10.1093/ptep/ptag096},
    url = {https://doi.org/10.1093/ptep/ptag096},
}

\end{document}